# Intrinsic Limitations of Single Layer Polychromatic Metalens for Virtual Reality Visors

**Myunghoo Lee[1], Ben Nguyen[1], Zhihao Zhou[1], Prannav Jayashanker[1,2], Shane Colburn[1], Ethan Tseng[4], Felix Heide[4], Praneeth Chakravarthula[3], Arka Majumdar[1,2], Johannes E. Fröch[1]**

[1]Department of Electrical Engineering, University of Washington, Seattle, WA 98195, USA.

[2]Department of Physics, University of Washington, Seattle, WA 98195, USA.

[3]Department of Computer Science, University of North Carolina at Chapel Hill, Chapel Hill, North Carolina 27599, USA.

[4]Department of Computer Science, Princeton University, Princeton, NJ 08540, USA.

corresponding author: jfroech@uw.edu

## Abstract

Virtual and augmented reality (VR/AR) visors require compact and lightweight optics. Metalenses have been widely proposed as ultrathin replacements for bulky refractive eyepieces, with performance typically assessed using point spread function (PSF) and modulation transfer function (MTF) measurements. Here, we design, fabricate, and characterize a single-layer silicon nitride metalens optimized for the three emission peaks of an RGB OLED display, and benchmark it against refractive and Fresnel eyepieces. Under coherent illumination, the metalens exhibits a tightly confined PSF and strong mid-to-high spatial-frequency MTF, suggesting excellent optical performance. However, when evaluated in a realistic system-level VR testbed incorporating incoherent OLED illumination, a dynamic-pupil eye model, and near-eye-relevant focal lengths, the same device exhibits pronounced ghosting and background haze. We show that these artifacts arise from the intrinsic multifocal nature of polychromatic diffractive focusing and demonstrate that common mitigation strategies such as narrowband filtering and long-focal-length relay optics merely mask, rather than resolve, the issue. Our results establish that meta-optics for AR/VR must be evaluated under realistic system-level conditions to reveal their true imaging performance.

## Introduction

Progress in Virtual Reality (VR) technologies has pushed boundaries over recent years enabling experiences that seemed futuristic a decade ago. However, we are reaching a critical inflection point as the optical architecture is becoming restrictive and often imposes a trade-off between resolution, field of view, and dynamic range [1], [2]. Although advancements in image quality have traditionally been accompanied by heavier and more complex optical systems, these put stringent limits on head ergonomics. The main challenge arises as multiple optical aberrations require correction, which necessitates complex lens assemblies with multiple elements, resulting in front-heavy headsets, which cause user fatigue and limit the duration of comfortable use [1], [2]. Further adoption of VR systems is thus limited as immersive experiences are restricted to short time spans.

To eliminate these restraints and reduce the form factor, Fresnel lenses have been proposed as a potential solution. These divide the continuous curvature of a refractive lens into concentric annular zones, while maintaining the local curvature of the lens. Unfortunately, the physical discontinuities at zone boundaries act as strong scattering centers, generating pronounced stray light and directional scattering artifacts, known as “God Rays” [3], [4]. As an alternative way to use thinner lenses while also reducing the physical track length, the industry has recently adopted folded optical architectures, most notably pancake lenses for VR and birdbath optics for augmented reality (AR). Pancake lenses utilize polarization-based optical folding, in which light is reflected multiple times in the system between half-mirrors and quarter-wave plates to effectively achieve a longer optical path length in a shorter track length, thus allowing thinner lenses. However, this architecture requires light to pass through polarization elements multiple times, resulting in a hard efficiency ceiling of 25% of the display’s initial light [5], [6], [7]. Furthermore, imperfect polarization rotation at high incidence angles introduces peripheral ghosting artifacts. Conversely, birdbath architectures, which utilize a beam splitter and a spherical mirror, enable optical see-through capabilities. Yet, they remain fundamentally constrained by the physical bulk required for their geometry and suffer from substantial light loss during the beam-splitting process, often necessitating artificially darkened environments to maintain image contrast.

Thus, to find better solutions for VR, new nanophotonic concepts have been recently explored. In particular, metasurfaces have received great interest, which are planar arrays of nanostructures that introduce a designed phase shift and/or control amplitude of the incident light on a subwavelength scale [8], [9], [10], [11]. Unlike classical lenses that focus light through refraction, metasurfaces manipulate amplitude, phase, and polarization of light based on the induced local amplitude/phase delay. By engineering their geometry, orientation, and material properties, it is possible to achieve complex optical functionalities enabling ultrathin, lightweight flat optics, with the potential to replace bulky refractive lenses. As such, metasurfaces have been successfully utilized for a wide array of advanced applications beyond simple focusing, including high-fidelity holography [12], [13], dynamic beam steering [14], [15], complex polarization control [16], [17], [18], and optical neural networks [19], [20].

However, transferring this concept to VR is facing significant challenges. A major obstacle has been extreme chromatic aberrations, whereas typical metalens designs, such as hyperboloid or quadratic phase profiles, result in a focal length that is inversely proportional to the light wavelength [21], [22]. Recent works in metalens have looked for solutions to this problem to provide broadband imaging capabilities [21], [23], [24], [25], [26], [27]. Yet, these solutions come with a trade-off in resolution, aperture size, numerical aperture (NA), and focus efficiency or require an additional computational imaging approach [28]. For instance, dispersion-engineered metalenses typically require complex nanostructure libraries to compensate for the differential group delays of various wavelengths, a requirement that becomes increasingly difficult to satisfy as the lens diameter increases [29].

Herein, we study the general applicability of inverse-designed polychromatic metalens for VR and reveal their intrinsic limitations. We utilized a previously developed inverse design framework that enables polychromatic focusing suitable for RGB bands [27]. Rather than evaluating metalens performance solely through isolated optical metrics, such as the point spread function, we further interpret their limitations through experimentally observed imaging artifacts that emerge in a VR testbed, under realistic eye-model considerations. We provide a baseline comparison with both refractive and Fresnel lenses to detail the origin of imaging artifacts, such as contrast loss, background haze, and ghosting. The direct physical insight connecting the mechanisms governing image formation in metalens yields a clearer understanding of the opportunities and limitations for VR. Based on these results we highlight points of action that need to be addressed in the future to make metalens suitable for compact and lightweight VR headsets.

## Results and Discussion

A typical virtual reality (VR) visor configuration is illustrated in Figure 1(a). Content is projected on a display, which is relayed through a lens assembly to the user's eye. To successfully transition these optical concepts into practical VR hardware, the lens design must satisfy several stringent system-level parameters simultaneously. Based on the geometric constraints of near-eye displays, achieving a comfortable user experience necessitates an eye relief, the physical clearance between the final optical element and the cornea, of at least 15 mm to 20 mm to adequately accommodate eyelashes and prescription eyewear [1], [30]. However, maintaining an extended eye relief while accommodating a wide angular field of view (FOV > 90°) fundamentally constrains the system's étendue, thereby requiring a highly expanded entrance pupil to prevent off-axis vignetting [31], [32], [33].

While the optical system must satisfy these parameters to provide the necessary FOV and user tolerance, doing so with conventional refractive lenses severely limits the wearability of the visor. Achieving aberration-free operation across such large apertures and wide field angles traditionally requires stacking multiple thick refractive elements. This significantly extends the total track length and introduces a heavy optical assembly, creating a downward torque on the user's head that accelerates fatigue [1], [34]. Consequently, the main advantage of metalens is highly compelling by compressing complex wavefront engineering into an ultrathin, planar surface. This provides a disruptive reduction in weight and form factor, directly alleviating torque-related ergonomic constraints [9], [35].

To provide a current overview of this rapidly advancing field, Table 1 details the primary characteristics of state-of-the-art meta-optical architectures. The table compares critical performance metrics, specifically aperture size, field of view, and chromatic behavior across a diverse range of approaches. These include millimeter-scale achromatic metalenses [36], inverse-designed large-area metalenses targeted for RGB operation [37], geometric phase metalens eyepieces [38], and recently demonstrated wide FOV doublets [39]. Collectively, these foundational studies highlight the potential of metalens to miniaturize near-eye visual systems.

***Table 1***. *Comparison of prior metalens VR systems.*

| **Optical approach** | **Representative work** | **Aperture / lens diameter (mm)** | **NA** | **Field of view** | **Chromatic behavior** | **System-level limitations for VR** |
|---|---|---|---|---|---|---|
| Refractive lens | Commercial VR optics | 10 | — | Narrow | Corrected via multi-element design | Bulky, heavy, long optical track length |
| Fresnel lens | Commercial VR optics | 10 | — | Wide | Residual chromatic aberration | Severe scattering, directional glare, reduced contrast |
| Achromatic meta-optics | Li *et al.* (2021) [36] | 2 | 0.7 | Narrow (<10°) | RGB (3λ) achromatic | Extremely limited eyebox, not VR-compatible |
| Inverse-designed meta-optics | Li *et al.* (2022) [37] | ~10 | 0.3 | — | RGB (3λ) achromatic | Low efficiency, used laser-based illumination |
| Geometric-phase metasurface eyepiece (AR) | Lee *et al.* (2018) [38] | 20 | 0.61* | 90° (mono) / 76° (full-color) | Strong chromatic focal shift | Requires system-level chromatic compensation |
| Wide FOV meta-optics doublet | Wirth-Singh *et al.* (2025) [39] | 21** | 0.18 | Wide (>60°) | Monochromatic | Single-wavelength operation, multi-layer architecture (Doublet) |
| Holographic meta-optics | Choi *et al.* (2025) [40] | 10 | 0.7*** | Wide (~60°) | RGB (3λ) achromatic | Low focusing efficiency (~ 10%), Relies on CGH to correct off-axis aberrations |
| **This work** | | 10 | 0.25 | Moderate (~ 30°) | RGB (3λ) achromatic (filtered) | Spectral sensitivity mitigated via filtering |

*NA of 0.61 corresponds to the red wavelength (660 nm); NA varies with wavelength (0.53 at 532 nm, 0.49 at 473 nm).
**Entrance aperture of the doublet (physical pupil 5.4 mm).
***Design NA; measured effective NA ≈ 0.5–0.55.

However, as we will point out, some of these optical metrics in isolation, such as PSF measurements, or MTF, may not directly relate to applicability, because certain constraints, such as pupil size and operating assumptions can differ substantially from those of a realistic VR display system and effects may only occur within specific pupil ranges. Some reported achromatic metalenses are demonstrated at apertures of only a few hundred micrometers or small NA, which severely restricts the eyebox and tolerable eye motion in VR use. Inverse-designed large-area metalenses partially address this limitation, but often do so at the expense of efficiency, or angular tolerance [37], [40]. Conversely, Lee et al. [38] demonstrated augmented reality configurations with emphasis on larger apertures and wide field of view yet typically rely on geometric-phase operation that introduces strong chromatic focal shifts and requires additional system-level compensation. As a result, although these works demonstrate feasibility, their experimental configurations remain partially decoupled from the constraints imposed by realistic VR frameworks.

In more detail, we consider a design suitable for 3 wavelengths, specifically for an RGB OLED display, with 3 distinctive emission peaks, centered at 617 nm, 521 nm, and 474 nm. We use an inverse design approach, where the volume under the modulation transfer function (MTF) is maximized as described in the Supplementary Information. The optimized design was then fabricated in SiN, described in the Methods section. The fabricated meta-optic compared with a refractive lens of the same parameters is shown in Figure 1(b). Optical and Scanning Electron Microscope images show the high fabrication quality of the optic. Compared to a refractive lens with the same focal length and aperture size, we yield a reduction of 10 x in thickness and weight, as highlighted in Figure 1(b).

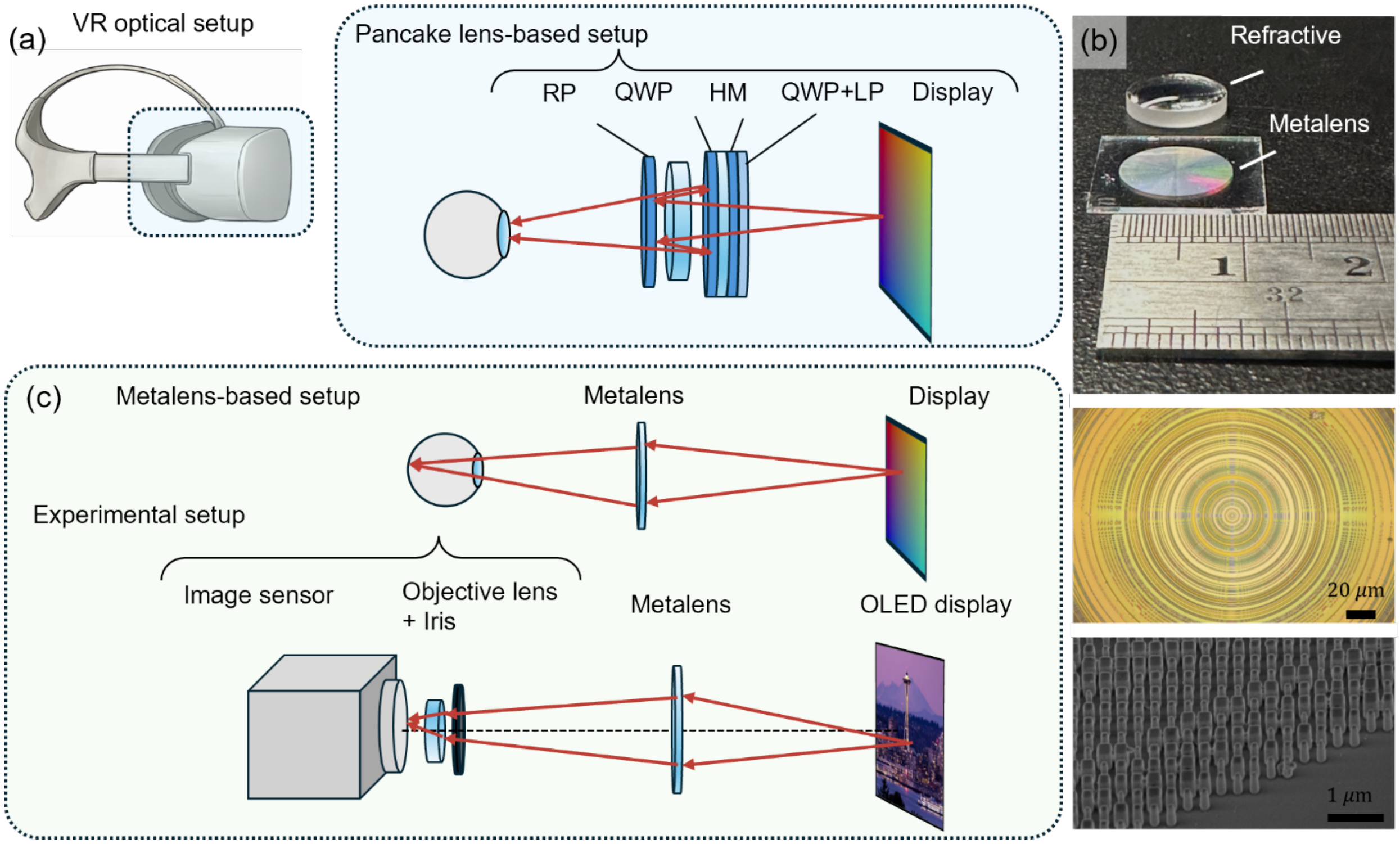

***Figure 1. Comparison of near-eye virtual reality (VR) architecture and multi-scale metalens characterization.*** *(a) Conventional VR optical configuration featuring a pancake lens-based setup, which achieves a compact form factor through optical folding via a reflective polarizer (RP), quarter-wave plates (QWP), a half mirror (HM), and a linear polarizer (LP). (b) Multi-scale visualization of the fabricated device. A macroscopic photograph comparing a traditional refractive lens with flat metalens, an optical microscope image showing the concentric phase zones, and a scanning electron microscope (SEM) image of the underlying nanostructure array. (c) Proposed metalens-based setup (top) utilizing a single nanophotonic element to replace bulk optics, and the corresponding benchtop experimental setup (bottom) where a camera sensor paired with a lens and iris simulates the human eye to evaluate imaging performance using an OLED display.*

To evaluate the intrinsic focusing performance of the refractive, Fresnel, and metalens eyepiece, we measured the point spread functions (PSFs) at the designed RGB wavelengths (474 nm, 521 nm, and 617 nm) using a laser-based setup and subsequently derived the corresponding modulation transfer functions (MTFs). In addition, we measured the focusing profile, using a programmable stage to reveal the intensity profile along the optical axis.

As shown in Figure 2, both the refractive lens and the meta-optic exhibit a tightly confined PSF with minimal side lobes. However, the derived MTF data reveals a more nuanced performance comparison across the spatial frequency spectrum. At the target focal plane, the meta-optic demonstrates a well-confined PSF that is considerably sharper than the Fresnel lens, exhibiting reduced background scattering.

While the conventional refractive lens maintains a higher MTF at low spatial frequencies, the meta-optic exhibits superior contrast transfer at mid-to-high spatial frequencies. Specifically, in the green band (521 nm), the MTF of the meta-optic overtakes that of the refractive lens in the spatial frequency range between approximately 70 lp/mm and 130 lp/mm. For the blue (474 nm) and red (617 nm) bands, the meta-optic consistently outperforms the refractive lens across the majority of the spectrum, maintaining a higher contrast from approximately 10 lp/mm up to 200 lp/mm. Furthermore, the meta-optic significantly outperforms the Fresnel lens across all measured wavelengths and spatial frequencies.

This enhanced mid-to-high frequency contrast transfer indicates that the subwavelength wavefront-engineered metasurface effectively mitigates certain higher-order aberrations inherently present in thick, single-element refractive lenses. These results confirm that under coherent laser illumination, the meta-optic achieves highly competitive focusing relative to traditional refractive optics and surpasses it within critical resolution bands, providing a definitive improvement over Fresnel-based architectures. In comparison, the Fresnel lens demonstrates significant scattering, an extended PSF, and pronounced background artifacts such as streaks, resulting in a broadened focal spot and substantially degraded MTF response, especially at mid-to-high spatial frequencies.

To further illustrate the origin of the polychromatic focusing capabilities, we measured the longitudinal depth profile (shown in Supplementary Figure S2). It is important to note that this multifocal behavior arises fundamentally from the polychromatic inverse design. Essentially, diffractive optics focus light according to the phase term $e^{ikz}$. However, the wavevector $kz$ introduces ambiguity, as enforcing a common focal plane for the three distinct design wavelengths (474 nm, 521 nm, and 617 nm) inevitably generates multiple, spatially separated focal spots for any single wavelength along the optical axis.

To understand the performance in a realistic VR configuration using an OLED display, we compared the refractive, Fresnel, and metalens eyepieces in imaging scenarios, as shown in Figure 3. We use a basic eye model consisting of an adjustable aperture (set to 5 mm) and refractive lens positioned in front of an image sensor (ASI183MC Pro, ZWO, Pixel size = 2.4 um). This setup allowed us to evaluate lens performance under conditions representative of commercial OLED-based VR systems. Unlike other studies on metalens for VR, which required laser-based coherent illumination or small aperture conditions, our experiments were performed using an incoherent OLED display and realistic aperture sizes and eye model, providing a more application-relevant evaluation framework.

First, we projected an array of white dots over the display area, (Fig. 3(a)), to understand the performance at varying field angles. While the refractive lens exhibits the sharpest and most contrasted dots near the center, the effective field of view is extremely limited; peripheral dots (yellow inset) are significantly attenuated and blurred, exhibiting strong coma, indicative of strong off-axis aberrations. The Fresnel lens exhibits reduced central sharpness compared to the refractive lens and, more critically, pronounced directional flare artifacts at large field angles. These streak-like features arise from discontinuities at the Fresnel zone boundaries, where non-ideal scattering for high-angle incidence rays causes flare.

The metalens eyepiece under broadband OLED illumination exhibits reduced image contrast, due to a noticeable chromatic blur across the field. Unlike the Fresnel lens, this degradation does not originate from structural discontinuities but rather from chromatic defocus introduced by non-designed wavelengths within the broadband spectrum, which is not accurately captured in monochromatic laser-based PSF measurements. This can be mitigated by introducing narrowband filters corresponding to the distinct optimized red, green, and blue wavelengths of the meta-optic. Experimentally, we evaluated the meta-optic performance using three separate bandpass filters with central wavelengths and full width at half-maximum (FWHM) bandwidths of 620 nm (FWHM = 10 nm), 520 nm (FWHM = 10 nm), and 473 nm (FWHM = 3 nm) (lower rows of Figure 3(a)). With the proper filters isolating the designed operational bands, the dot contrast improves substantially, and field uniformity is restored even at larger field angles. This confirms that these aberrations primarily arise due to residual spectral components rather than fundamental wavefront errors.

However, we also observe that a residual background haze remains, impacting the contrast over the entire field. In further experiments (Fig. 3(b)) on a more extended scene, we observe that this haziness takes shape with respect to the original image, and a somewhat blurred ghost image appears. Moreover, because these features are not congruent with the intrinsic periodic arrangement of the nanopillars (arranged in a square lattice), we exclude higher order diffraction related effects arising from the periodic pillar arrangement. As the ghosting features appear with directionality towards the optical axis, it points towards a diffraction related effect stemming from the phase profile.

The origin of this effect is linked to the multiple focal spots of the metalens. Although these enable polychromatic focusing, it causes in reverse a single point source to produce three distinct wavefront patterns in the far-field. A portion of the light will be collimated, while others are not. To validate our hypothesis that the observed ghost effect and background haze originate from the intrinsic multi-focusing behavior of the meta-optic, we systematically modified the collection optics in our experimental testbed, as detailed in Figure 4. By adjusting the aperture size of the iris, which simulates the human pupil, we directly manipulated the depth of field of the imaging system. When the aperture size is reduced (Fig. 4(a)), the depth of field increases. Diverging or converging wavefronts corresponding to non-target wavelengths that are not perfectly collimated by the meta-optic are captured with sufficient sharpness to form distinct secondary images on the sensor. This is specifically well indicated by the very different formation of ghost images that are observed in experiments for red, green, and blue. In comparison, when the aperture is widened, as shown in Figure 4(b), the depth of field becomes shallower, causing these secondary focal planes to blur into a diffuse background haze rather than structured ghost images.

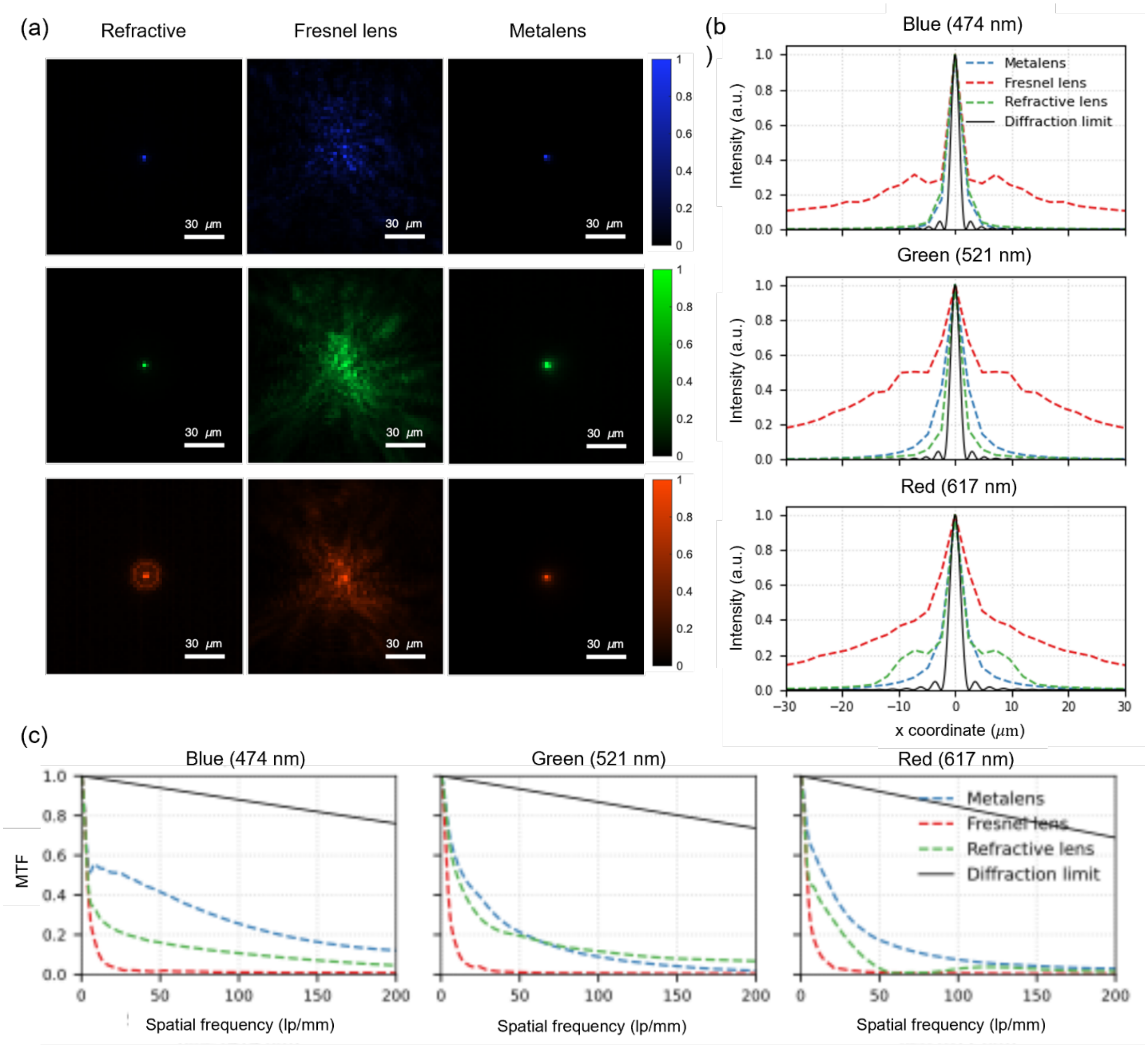

***Figure 2. Focusing performance of refractive, Fresnel, and meta-optic eyepieces.*** *(a) Measured point spread functions (PSFs) and corresponding modulation transfer functions (MTFs) at designed RGB wavelengths (474 nm, 521 nm, and 617 nm – top to bottom). (b) 1D PSF comparison between diffraction limit, refractive, Fresnel, and metalens at designed RGB wavelengths. (c) Corresponding modulation transfer functions (MTFs) comparison.*

To further test the influence of the evaluation optical setup, we replaced the eye model with a standard tube lens configuration (Fig. 4(c)). Under these conditions, the captured image appears significantly sharper, and the observed imaging artifacts completely vanish. This discrepancy highlights a fundamental vulnerability in evaluating metalens for display applications. The multi-image phenomenon is a direct consequence of the diffractive nature of the metalens, governed by the inherent relationship $f \times \lambda = const$, where $f$ is the focal length and $\lambda$ represents the wavelength of the incident light [22], [41]. For a given phase profile, establishing a primary focal point for the red band inevitably generates secondary, spatially separated focal planes for the green and blue bands. However, with a long tube lens focal length, the non-desired wavefront components are strongly diluted across the image plane, causing them to fall below the detector noise to be captured. Thus, the image appears most crisp under this non-realistic configuration for red and blue wavelengths [36], [37], [40].

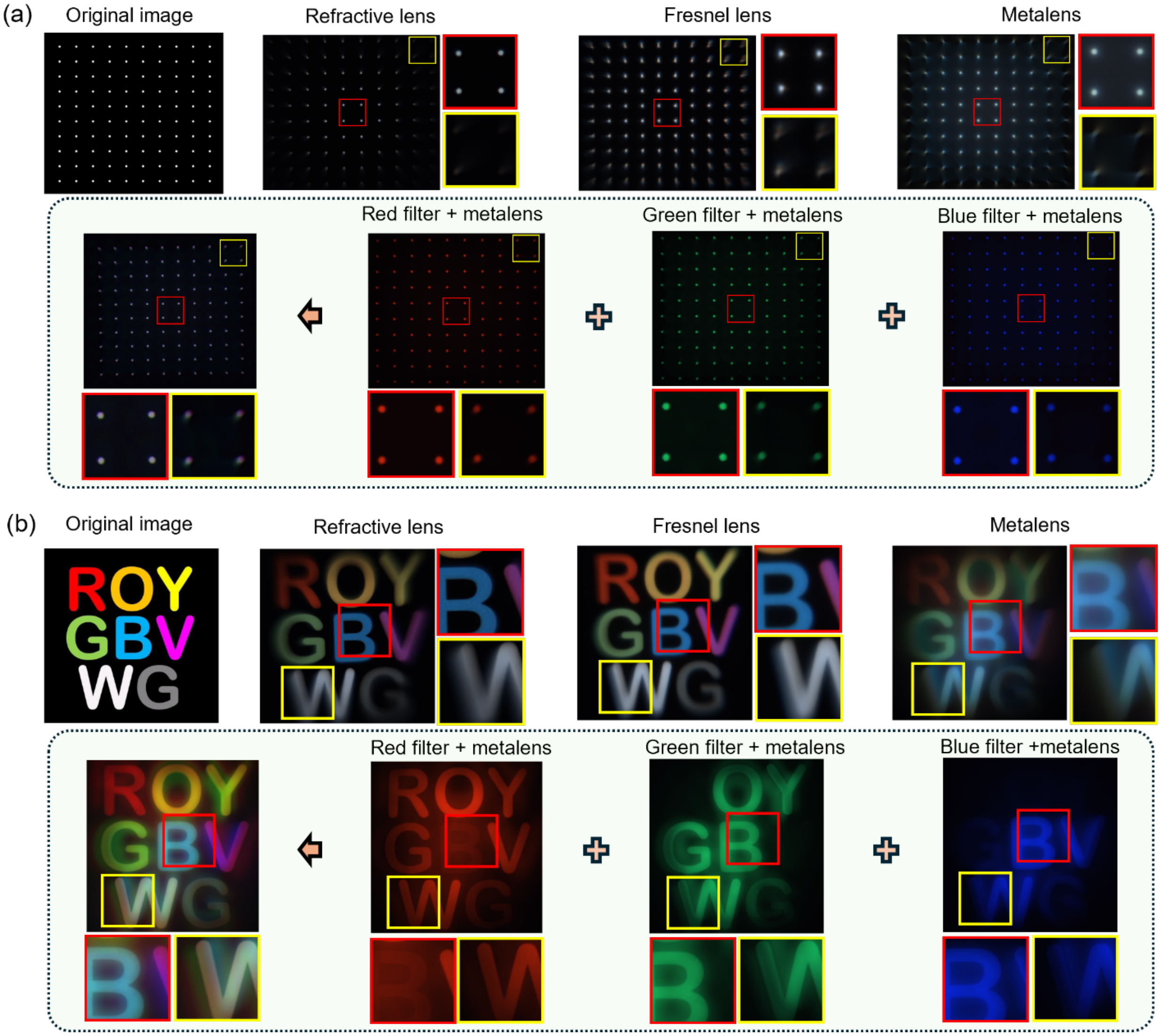

***Figure 3. Imaging quality and aberrations of conventional lenses and metalens.*** *(a) Imaging results of a white dot array pattern. The top row displays images captured under broadband illumination using a standard refractive lens, a Fresnel lens, and the metalens. The subsequent rows show the metalens performance under narrowband illumination using red, green, and blue filters and the combined image from those three images. (b) Imaging results of a multi-colored alphanumeric test pattern ("ROY GBV WG"). The layout follows the same arrangement as in (a).*

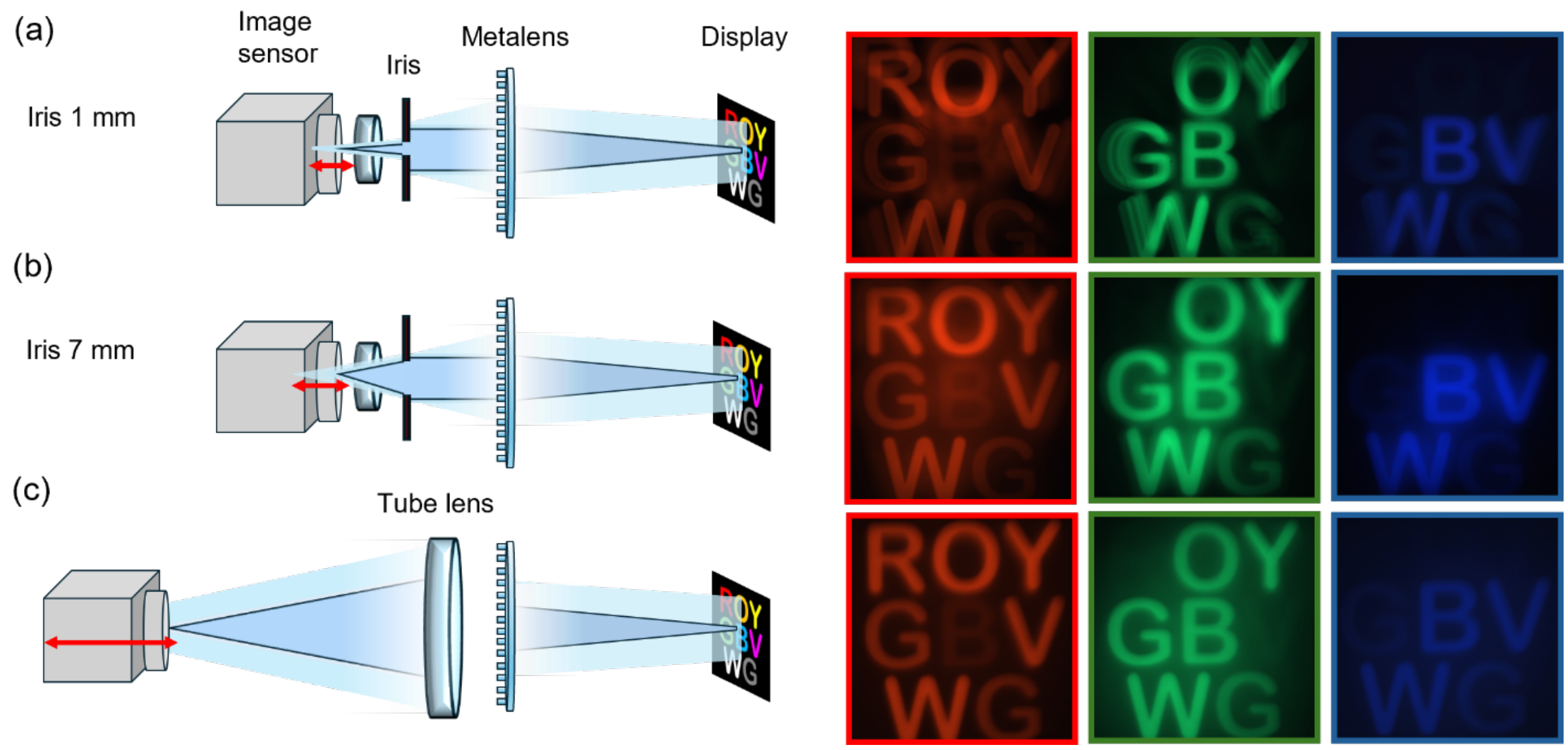

***Figure 4. Impact of aperture size and collection optics on multi-focusing artifacts.*** *(a) Small iris aperture increases depth of field, revealing distinct secondary images from residual wavelengths. (b) Large iris aperture decreases depth of field, blurring secondary images into background haze. (c) Standard tube lens configuration artificially suppresses these artifacts, highlighting the discrepancy with realistic near-eye evaluations.*

To further evaluate the impact of these intrinsic characteristics within a realistic near-eye display architecture, we further examined image formation using more complex scenes, as shown in Figure 5. Compared with structured test targets, natural images contain continuous content mixed with dense spatial frequencies, providing a more stringent assessment of practical VR operation. The refractive eyepiece maintains relatively high central sharpness; however, noticeable field-dependent degradation and peripheral blur are observed, consistent with its limited effective field of view. In contrast, the metalens eyepiece, when operated at the designed RGB bands, preserves image structure across a larger portion of the field without exhibiting the flare-like scattering artifacts characteristic of Fresnel zone discontinuities. This indicates that the continuous phase profile of the metasurface enables controlled wavefront shaping within a planar geometry while mitigating high-angle structural scattering. However, multifocal behavior also effectively causes blurriness within realistic VR settings as can be seen across various images. Importantly this example shows that ghost images are less identifiable when complex scenes are projected on display. We note that color accuracy, including hue and saturation can be adjusted/ calibrated when the image is projected.

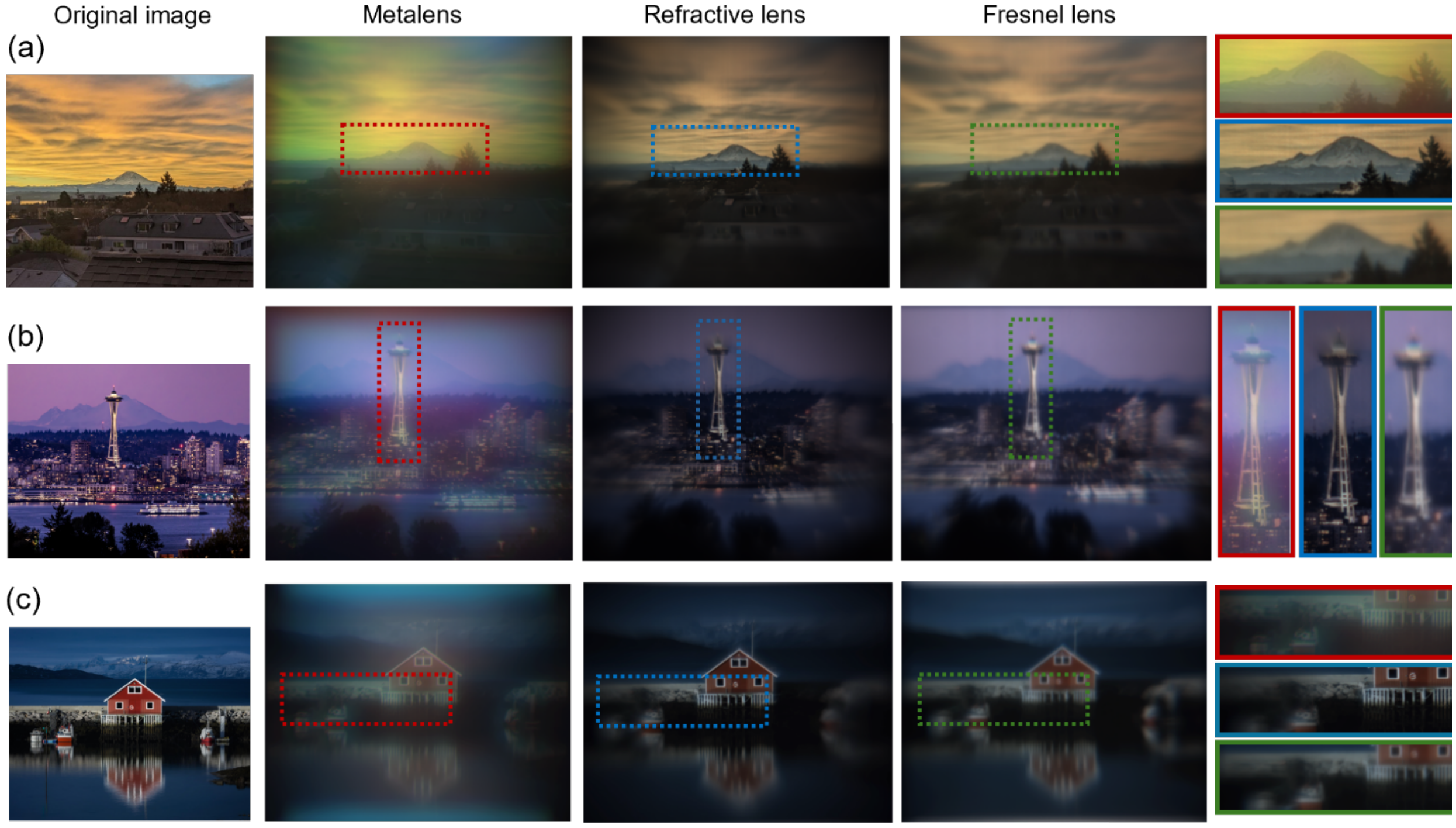

***Figure 5. Imaging assessment for complex scenes.*** *(a-c) Experimental results comparing the original image with captures from the metalens, a refractive lens, and a Fresnel lens under simulated virtual reality conditions. Magnified insets detail differences in sharpness, contrast, and edge fidelity.*

## Conclusion

This study exposes a critical vulnerability in the current evaluation paradigms of meta-optics for AR/VR applications. We demonstrate that conventional metrics, such as isolated PSF and MTF, are fundamentally insufficient and often deceptive when decoupled from system-level architectures. While these metrics provide valuable insight into wave-optical behavior, they may not necessarily correlate with perceived image quality in practical VR scenarios. Specifically, utilizing standard relay optics or tube lenses inadvertently masks severe multi-focusing artifacts, yielding an overly optimistic assessment of full-color display performance that fails to translate to near-eye environments. Therefore, it is imperative that future meta-optics for compact AR/VR visors be evaluated strictly under realistic, system-level viewing conditions without corrective optics, meticulously accounting for dynamic pupil constraints ranging from 2 to 8 mm.

By establishing a system-level experimental platform that mimics realistic operating conditions, we clarify both the strengths such as compactness and wavefront engineering flexibility and the limitations such as chromatic sensitivity and system trade-offs of polychromatic metasurfaces. Addressing focal separation represents the next critical steps in advancing metalens eyepieces toward fully optimized practical VR systems.

We demonstrate that narrowband filters can effectively mitigate the chromatic contrast degradation. However, we also point out that a more severe limitation stems from the inherent dependence - between wavelength and focal length of a diffractive optics. Breaking this dependence remains an open problem. We envision several potential pathways that could address this problem. First, would be the design of a meta-optic that achieves polychromatic focusing through harmonic Fresnel zone engineering. This will require the total phase of several wavelengths to be a common multiple of the same least value [41], [42], [43], [44]. However, a major limitation in this case becomes clear as it will require pillars with extreme aspect ratios, as high phase shifts will need to be implemented. Another approach could be by designing additional layers of non-local metasurfaces, that locally filter light based on the angular properties of the incoming light.[45], [46]

Ultimately, this study serves as a necessary recalibration for the field, bridging the disconnect between idealized laboratory demonstrations and deployable VR hardware. By offering a rigorously grounded assessment, we define the practical boundaries of current metalens technologies and outline the necessary physical and engineering milestones for next-generation near-eye displays.

## Methods

### Design and Fabrication of the polychromatic metalens

To design the polychromatic meta-optic, we adapted the inverse design framework previously reported for broadband nano-optics [27]. This computational approach utilizes a gradient-based optimization technique to iteratively refine the spatial distribution of the metasurface, maximizing the volume under the modulation transfer function (MTF) to ensure high-fidelity contrast transfer. While the original framework optimized performance over a continuous spectral bandwidth, our optimization space was strategically restricted to three discrete target wavelengths (474 nm, 521 nm, and 617 nm) to directly match the primary emission peaks of the RGB OLED display. By focusing the inverse design specifically on these discrete bands, we derived an optimal phase profile that maximizes the MTF volume exclusively for the targeted wavelengths, ensuring efficient polychromatic focusing.

The optimized metasurface profile was physically realized as a silicon nitride (SiN) nanostructure array on a fused silica substrate using standard nanofabrication procedures. Initially, a SiN thin film was deposited on the quartz substrate via plasma-enhanced chemical vapor deposition (PECVD). A high-resolution positive electron-beam resist (ZEP-520A) was then spin-coated onto the substrate, and the optimized layout was patterned via electron-beam lithography (EBL) followed by development. For precise pattern transfer, an aluminum oxide ($Al_2O_3$) hard mask layer was deposited using an electron-beam evaporator, and the mask layout was defined through a subsequent lift-off process. Finally, the nanostructure pattern was anisotropically transferred into the underlying SiN layer via inductively coupled plasma (ICP) reactive ion etching, yielding the final single-layer polychromatic meta-optic.

### Optical Measurements – PSF

To characterize the intrinsic focusing performance and point spread function (PSF) of the optical elements, we utilized a custom-built laser-based measurement setup. Illumination was provided by a broadband supercontinuum laser source (SUPERK Fianium FIR-20, NKT Photonics). To isolate the specific design wavelengths of the polychromatic meta-optic (474 nm, 521 nm, and 617 nm), the broadband emission was directed through an acousto-optic tunable filter (AOTF, SuperK Select, NKT Photonics). The spectrally filtered light was subsequently coupled into a single-mode fiber (P1-630Y-FC-2, Thorlabs). The output facet of this fiber was positioned approximately 50 cm away from the optical element under test to approximate a collimated incident wavefront.

The transmitted light focused by the optics was directly captured using a high-resolution image sensor (ASI183MC Pro, ZWO, 2.4 µm pixel pitch). To measure the longitudinal depth profile and characterize the multifocal behavior along the optical axis, the image sensor was mounted on a programmable motorized linear translation stage. This configuration enabled precise spatial scanning of the focal volume to extract the intensity distribution along the longitudinal direction. To derive the modulation transfer functions (MTFs) while mitigating spectral leakage, the captured two-dimensional PSF images were further processed. First, the sensor background noise was subtracted using the median intensity of the non-illuminated edge pixels. A spatial Tukey window was then applied to the cropped PSF to attenuate the boundary regions to zero, suppressing high-frequency artifacts during the Fourier transform. The windowed data was subsequently zero-padded to increase the frequency resolution before applying a two-dimension fast Fourier transform (FFT). Finally, the resulting two-dimension MTF was radially averaged to extract the one-dimension MTF profile.

**Optical Measurements – VR configuration**

To evaluate the imaging performance of the optical elements under realistic near-eye display conditions, we constructed a benchtop virtual reality testbed. A 0.7-inch Micro OLED display (1920 × 1080 resolution, capable of a maximum brightness of 3000 nits) was utilized as the incoherent image source to project the structured test patterns and complex full-color scenes. The human visual system was approximated using a custom-built eye model comprising a high-resolution image sensor (ASI183MC Pro, ZWO) paired with an adjustable iris to simulate the human pupil. An uncoated N-BK7 plano-convex lens (LA1560, Thorlabs, f = 25 mm, Ø1/2") was mounted directly in front of the sensor using standard optomechanical spacers to represent the crystalline lens.

For the comparative imaging assessment, the meta-optic (10 mm aperture, f = 20 mm) was benchmarked against a conventional refractive lens and a Fresnel lens. The refractive reference was an uncoated plano-convex lens (Edmund Optics) with an identical 10 mm diameter and 20 mm focal length. The Fresnel lens reference (FRP125, Thorlabs, f = 25 mm, Ø1") was fitted with an external iris to restrict its clear aperture to 12.5 mm. This precise aperture restriction ensured a consistent focal ratio of f/2 across the meta-optic, the refractive lens, and the Fresnel lens, thereby establishing a rigorous and fair evaluation of their intrinsic optical performance.

To further investigate the multi-focusing artifacts and the influence of the evaluation optics on the perceived image quality, we performed additional verifications by systematically modifying the collection system. In these specific experiments, the simplified eye model was temporarily substituted with a standard tube lens configuration. A broadband $MgF_2$-coated tube lens (TTL200-S8, Thorlabs, f = 200 mm) was employed to relay the transmitted light directly onto the sensor. This configuration was utilized to demonstrate how conventional optical evaluation methods, which rely on standard tube lenses, can inadvertently mask the severe multi-focusing artifacts that emerge under realistic near-eye viewing conditions.

**Data Availability**

Data available on request from the authors.

**Acknowledgements**

Part of this work was conducted at the Washington Nanofabrication Facility / Molecular Analysis Facility, a National Nanotechnology Coordinated Infrastructure (NNCI) site at the University of Washington with partial support from the National Science Foundation via awards NNCI-1542101 and NNCI-2025489.

**References**

[1] B. C. Kress, "Optical Architectures for Augmented-, Virtual-, and Mixed-Reality Headsets".
[2] D. M. Hoffman, A. R. Girshick, K. Akeley, and M. S. Banks, "Vergence–accommodation conflicts hinder visual performance and cause visual fatigue," *J. Vis.*, vol. 8, no. 3, pp. 33–33, 2008.
[3] D. Cheng *et al.*, "Optical design and pupil swim analysis of a compact, large EPD and immersive VR head mounted display," *Opt. Express*, vol. 30, no. 5, pp. 6584–6602, 2022.
[4] S. Grogorick, M. Ueberheide, J.-P. Tauscher, P. M. Bittner, and M. Magnor, "Gaze and motion-aware real-time dome projection system," in *2019 IEEE Conference on Virtual Reality and 3D User Interfaces (VR)*, IEEE, 2019, pp. 1780–1783. Accessed: Jun. 04, 2026. [Online]. Available: https://ieeexplore.ieee.org/abstract/document/8797902/?casa_token=rnXUKUeiUfEAAAAA:vEeH39_LSsPfUemK2EwAZ7lobwKIMBDqJEMWQuRjk7hxOirltBdCH-r5Pz_29al9LR24ZTrdCQ
[5] Q. Hou *et al.*, "Stray light analysis and suppression method of a pancake virtual reality head-mounted display," *Opt. Express*, vol. 30, no. 25, pp. 44918–44932, 2022.
[6] Z. Luo, Y. Ding, Q. Yang, and S.-T. Wu, "Ghost image analysis for pancake virtual reality systems," *Opt. Express*, vol. 32, no. 10, pp. 17211–17219, 2024.
[7] Y. Ding, Z. Luo, G. Borjigin, and S.-T. Wu, "Breaking the optical efficiency limit of virtual reality with a nonreciprocal polarization rotator," *Opto-Electron. Adv.*, vol. 7, no. 3, pp. 230178–1, 2024.
[8] N. Yu and F. Capasso, "Flat optics with designer metasurfaces," *Nat. Mater.*, vol. 13, no. 2, pp. 139–150, 2014.
[9] A. V. Kildishev, A. Boltasseva, and V. M. Shalaev, "Planar Photonics with Metasurfaces," *Science*, vol. 339, no. 6125, p. 1232009, Mar. 2013, doi: 10.1126/science.1232009.
[10] D. Lin, P. Fan, E. Hasman, and M. L. Brongersma, "Dielectric gradient metasurface optical elements," *Science*, vol. 345, no. 6194, pp. 298–302, Jul. 2014, doi: 10.1126/science.1253213.
[11] M. Khorasaninejad, W. T. Chen, R. C. Devlin, J. Oh, A. Y. Zhu, and F. Capasso, "Metalenses at visible wavelengths: Diffraction-limited focusing and subwavelength resolution imaging," *Science*, vol. 352, no. 6290, pp. 1190–1194, Jun. 2016, doi: 10.1126/science.aaf6644.

[12] G. Zheng, H. Mühlenbernd, M. Kenney, G. Li, T. Zentgraf, and S. Zhang, “Metasurface holograms reaching 80% efficiency,” *Nat. Nanotechnol.*, vol. 10, no. 4, pp. 308–312, 2015.

[13] L. Huang *et al.*, “Three-dimensional optical holography using a plasmonic metasurface,” *Nat. Commun.*, vol. 4, no. 1, p. 2808, 2013.

[14] N. Yu *et al.*, “Light Propagation with Phase Discontinuities: Generalized Laws of Reflection and Refraction,” *Science*, vol. 334, no. 6054, pp. 333–337, Oct. 2011, doi: 10.1126/science.1210713.

[15] G. K. Shirmanesh, R. Sokhoyan, P. C. Wu, and H. A. Atwater, “Electro-optically Tunable Multifunctional Metasurfaces,” *ACS Nano*, vol. 14, no. 6, pp. 6912–6920, Jun. 2020, doi: 10.1021/acsnano.0c01269.

[16] A. Arbabi, Y. Horie, M. Bagheri, and A. Faraon, “Dielectric metasurfaces for complete control of phase and polarization with subwavelength spatial resolution and high transmission,” *Nat. Nanotechnol.*, vol. 10, no. 11, pp. 937–943, 2015.

[17] N. A. Rubin, G. D’Aversa, P. Chevalier, Z. Shi, W. T. Chen, and F. Capasso, “Matrix Fourier optics enables a compact full-Stokes polarization camera,” *Science*, vol. 365, no. 6448, p. eaax1839, Jul. 2019, doi: 10.1126/science.aax1839.

[18] J. P. Balthasar Mueller, N. A. Rubin, R. C. Devlin, B. Groever, and F. Capasso, “Metasurface Polarization Optics: Independent Phase Control of Arbitrary Orthogonal States of Polarization,” *Phys. Rev. Lett.*, vol. 118, no. 11, p. 113901, Mar. 2017, doi: 10.1103/PhysRevLett.118.113901.

[19] X. Lin *et al.*, “All-optical machine learning using diffractive deep neural networks,” *Science*, vol. 361, no. 6406, pp. 1004–1008, Sep. 2018, doi: 10.1126/science.aat8084.

[20] Y. Luo *et al.*, “Design of task-specific optical systems using broadband diffractive neural networks,” *Light Sci. Appl.*, vol. 8, no. 1, p. 112, 2019.

[21] S. Wang *et al.*, “A broadband achromatic metalens in the visible,” *Nat. Nanotechnol.*, vol. 13, no. 3, pp. 227–232, 2018.

[22] E. Arbabi, A. Arbabi, S. M. Kamali, Y. Horie, and A. Faraon, “Multiwavelength polarization-insensitive lenses based on dielectric metasurfaces with meta-molecules,” *Optica*, vol. 3, no. 6, p. 628, Jun. 2016, doi: 10.1364/OPTICA.3.000628.

[23] F. Aieta, M. A. Kats, P. Genevet, and F. Capasso, “Multiwavelength achromatic metasurfaces by dispersive phase compensation,” *Science*, vol. 347, no. 6228, pp. 1342–1345, Mar. 2015, doi: 10.1126/science.aaa2494.

[24] W. T. Chen *et al.*, “A broadband achromatic metalens for focusing and imaging in the visible,” *Nat. Nanotechnol.*, vol. 13, no. 3, pp. 220–226, 2018.

[25] S. Shrestha, A. C. Overvig, M. Lu, A. Stein, and N. Yu, “Broadband achromatic dielectric metalenses,” *Light Sci. Appl.*, vol. 7, no. 1, p. 85, 2018.

[26] S. Colburn, A. Zhan, and A. Majumdar, “Metasurface optics for full-color computational imaging,” *Sci. Adv.*, vol. 4, no. 2, p. eaar2114, Feb. 2018, doi: 10.1126/sciadv.aar2114.

[27] J. E. Fröch *et al.*, “Beating spectral bandwidth limits for large aperture broadband nano-optics,” *Nat. Commun.*, vol. 16, no. 1, p. 3025, 2025.

[28] F. Presutti and F. Monticone, “Focusing on bandwidth: achromatic metalens limits,” *Optica*, vol. 7, no. 6, pp. 624–631, 2020.

[29] W. T. Chen, A. Y. Zhu, and F. Capasso, “Flat optics with dispersion-engineered metasurfaces,” *Nat. Rev. Mater.*, vol. 5, no. 8, pp. 604–620, 2020.

[30] S. A. Cholewiak, Z. Başgöze, O. Cakmakci, D. M. Hoffman, and E. A. Cooper, “A perceptual eyebox for near-eye displays,” *Opt. Express*, vol. 28, no. 25, pp. 38008–38028, 2020.

[31] T. Zhan, K. Yin, J. Xiong, Z. He, and S.-T. Wu, “Augmented reality and virtual reality displays: perspectives and challenges,” *Iscience*, vol. 23, no. 8, 2020, Accessed: Jun. 04, 2026. [Online]. Available: https://www.cell.com/iscience/fulltext/S2589-0042(20)30585-X

[32] D. Lanman and D. Luebke, “Near-eye light field displays,” *ACM Trans. Graph.*, vol. 32, no. 6, pp. 1–10, Nov. 2013, doi: 10.1145/2508363.2508366.

[33] B. Chao, M. Gopakumar, S. Choi, J. Kim, L. Shi, and G. Wetzstein, “Large Étendue 3D Holographic Display with Content-adaptive Dynamic Fourier Modulation,” in *SIGGRAPH Asia 2024 Conference Papers*, Tokyo Japan: ACM, Dec. 2024, pp. 1–12. doi: 10.1145/3680528.3687600.

[34] H. L. Gallagher, E. Caldwell, and C. B. Albery, “Neck muscle fatigue resulting from prolonged wear of weighted helmets,” 2008. Accessed: Jun. 04, 2026. [Online]. Available: https://apps.dtic.mil/sti/html/tr/ADA491626/

[35] J. Engelberg and U. Levy, “The advantages of metalenses over diffractive lenses,” *Nat. Commun.*, vol. 11, no. 1, p. 1991, 2020.

[36] Z. Li *et al.*, “Meta-optics achieves RGB-achromatic focusing for virtual reality,” *Sci. Adv.*, vol. 7, no. 5, p. eabe4458, Jan. 2021, doi: 10.1126/sciadv.abe4458.

[37] Z. Li, R. Pestourie, J.-S. Park, Y.-W. Huang, S. G. Johnson, and F. Capasso, “Inverse design enables large-scale high-performance meta-optics reshaping virtual reality,” *Nat. Commun.*, vol. 13, no. 1, p. 2409, 2022.

[38] G.-Y. Lee *et al.*, “Metasurface eyepiece for augmented reality,” *Nat. Commun.*, vol. 9, no. 1, p. 4562, 2018.

[39] A. Wirth-Singh *et al.*, “Wide field of view large aperture meta-doublet eyepiece,” *Light Sci. Appl.*, vol. 14, no. 1, p. 17, 2025.

[40] M. Choi *et al.*, “Roll-to-plate printable RGB achromatic metalens for wide-field-of-view holographic near-eye displays,” *Nat. Mater.*, vol. 24, no. 4, pp. 535–543, Apr. 2025, doi: 10.1038/s41563-025-02121-0.

[41] D. W. Sweeney and G. E. Sommargren, “Harmonic diffractive lenses,” *Appl. Opt.*, vol. 34, no. 14, pp. 2469–2475, 1995.

[42] D. Faklis and G. M. Morris, “Spectral properties of multiorder diffractive lenses,” *Appl. Opt.*, vol. 34, no. 14, pp. 2462–2468, 1995.

[43] Y. Hongli, C. Zhaofeng, and L. Xiaotong, “Achromatic and wide field of view metalens based on the harmonic diffraction and a quadratic phase,” *Opt. Express*, vol. 30, no. 25, pp. 45413–45425, Dec. 2022, doi: 10.1364/OE.475337.

[44] Q. Chen, Y. Gao, S. Pian, and Y. Ma, “Theory and Fundamental Limit of Quasiachromatic Metalens by Phase Delay Extension,” *Phys. Rev. Lett.*, vol. 131, no. 19, p. 193801, Nov. 2023, doi: 10.1103/PhysRevLett.131.193801.

[45] O. Reshef *et al.*, “An optic to replace space and its application towards ultra-thin imaging systems,” *Nat. Commun.*, vol. 12, no. 1, p. 3512, 2021.

[46] H. Kwon, D. Sounas, A. Cordaro, A. Polman, and A. Alù, “Nonlocal Metasurfaces for Optical Signal Processing,” *Phys. Rev. Lett.*, vol. 121, no. 17, p. 173004, Oct. 2018, doi: 10.1103/PhysRevLett.121.173004.